\documentclass[runningheads,a4paper]{llncs}

\usepackage{amssymb}
\usepackage{graphicx}

\usepackage{url}
\newcommand{\keywords}[1]{\par\addvspace\baselineskip
\noindent\keywordname\enspace\ignorespaces#1}

\begin{document}

\mainmatter  

\title{Quantum fully homomorphic encryption scheme based on universal quantum circuit}

\titlerunning{Quantum fully homomorphic encryption}

%

\author{Min Liang}

\authorrunning{Min Liang}

\institute{Data Communication Science and Technology Research Institute, \\Beijing 100191, China \\
liangmin07@mails.ucas.ac.cn}

%
%

\maketitle

\vspace{-5mm}
\begin{abstract}
Fully homomorphic encryption enables arbitrary computation on encrypted data without decrypting the data. Here it is studied in the context of quantum information processing. Based on universal quantum circuit, we present a quantum fully homomorphic encryption (QFHE) scheme, which permits arbitrary quantum transformation on an encrypted data. The QFHE scheme is proved to be perfectly secure. In the scheme, the decryption key is different from the encryption key, however, the encryption key cannot be public. Moreover, the evaluate algorithm of the scheme is independent of the encryption key, so it is very applicable in delegated quantum computing between two parties.
\end{abstract}

\vspace{-8mm}
\section{Introduction}
Suppose you have some encrypted data, is it possible to compute on the encrypted data without decrypting them?

This problem is relative to the research of ``delegated computing''. It has been investigated a lot in modern cryptography, such as homomorphic encryption \cite{rivest1978} and blind computing \cite{feigenbaum1986,abadi1989}. It was firstly considered by Rivest et al. who suggested some homomorphic encryption schemes \cite{rivest1978}. However, these schemes are insecure \cite{brickell1987}. Later, the fully homomorphic encryption (FHE) schemes \cite{gentry2009,brakerski2011,brakerski2012} are proposed. These schemes are constructed based on some hard computational problems, and their security relies on the computational difficulty of these problems.

Is it possible to solve the above question in the context of quantum information? Lots of researches have provided a positive answer in blind quantum computation \cite{childs2005,arrighi2006,aharonov2010,broadbent2009,sueki2013,vedral2012,morimae2012a,morimae2012b,fitzsimons2012,giovannetti2013}. Blind quantum computation is a secure cloud quantum computing protocol which enables a client to delegate her quantum computation to a server without leaking anything about her data, algorithm and result. There are some optimization design \cite{mantri2013,li2014} and experimental researches \cite{barz2012,barz2013}.

This article considers the problem of homomorphic encryption in the context of quantum information processing: suppose arbitrary quantum plaintext $\sigma$ has been encrypted, can you perform any quantum operator $U$ directly on the ciphertext (without decrypting the ciphertext) and the obtained state can be decrypted and turned into the desired state $U\sigma U^\dagger$?

Rohde et al. \cite{rohde2012} studied quantum walk with encrypted data, and proposed a limited quantum homomorphic encryption (QHE) scheme using the Boson sampling and multi-walker quantum walk models. This QHE scheme is applicable in delegated computing of quantum walk.

Ref. \cite{liang2013a} presented the definitions of QHE and quantum fully homomorphic encryption (QFHE), and constructed a symmetric QFHE scheme. However, in the scheme, the evaluate algorithm is related to the secret key, so it is not suitable for secure delegated quantum computing.

In Ref. \cite{liang2013b}, we proposed a tripartite blind quantum computation (TBQC) scheme based on universal quantum circuit (UQC). Inspired by the TBQC scheme, we study the theory of QHE and construct a QFHE scheme based on UQC. In the scheme, the encryption key and decryption key are different, however, both of them should be kept secret. The scheme is applicable in delegated quantum computing, since the evaluate algorithm is unrelated to the secret key.

\section{Preliminaries}
In this section, we firstly introduce the UQC \cite{bera2010,liang2011}, quantum one-time pad (QOTP) \cite{boykin2003,boykin2002}, and then present a new definition of QHE.

\begin{figure}[htp!]
\begin{center}
\includegraphics[width=6cm]{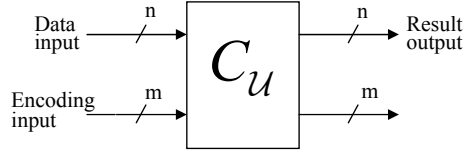}
\end{center}
\vspace{-5mm}
\caption{\label{fig1}An illustration of universal quantum circuit. The number ``n" indicates that the line represents a bunch of $n$ lines. The ancillary qubits can be seen as a part of the encoding, so it can be omitted here. In the circuit, the $m$-qubit output is independent of the $n$-qubit result.}
\end{figure}

{\bf Definition 1 (UQC \cite{bera2010})}. Fix $n>0$ and let $\mathcal{U}$ be a collection of unitary transformations on $n$ qubits. A quantum circuit $C_{\mathcal{U}}$ on $n+m$ qubits is universal for $\mathcal{U}$ if, for each transformation $U\in\mathcal{U}$, there is a string $e_U\in\{0,1\}^m$ (the encoding) such that for all strings $d\in\{0,1\}^n$ (the data),
\begin{equation}
C_\mathcal{U}(|d\rangle\otimes|e_U\rangle)=(U|d\rangle)\otimes|e_U\rangle.
\end{equation}
From the above definition, the UQC can be expressed as Figure \ref{fig1}.

In the UQC, given a $n$ qubits $d\rangle$ as the input data, when a $m$ qubits $|e_U\rangle$ is input as the encoding of a quantum transformation $U\in\mathcal{U}$, the UQC would output $n+m$ qubits $C_\mathcal{U}(|d\rangle\otimes|e_U\rangle)=(U|d\rangle)\otimes|e_U\rangle$. Here, $|e_U\rangle$ is called the encoding of the quantum transformation $U\in\mathcal{U}$ with regard to the UQC $C_\mathcal{U}$.

If there exists a UQC $C_\mathcal{U}$ for a family of quantum transformations $\mathcal{U}$, then each quantum transformation $U\in\mathcal{U}$ has a corresponding encoding $|e_U\rangle$, which satisfies the relation $C_\mathcal{U}(|d\rangle\otimes|e_U\rangle)=(U|d\rangle)\otimes|e_U\rangle$. If you want to carry out a series of quantum transformations in $\mathcal{U}$, you only need to know the encodings of these quantum transformations, and then perform the UQC $C_\mathcal{U}$ once for each encoding.

{\bf Definition 2 (QOTP \cite{boykin2003})}. Quantum one-time pad is a symmetric quantum encryption scheme. It is defined by a set $\{\frac{1}{2^{2n}},X^xZ^z|x,z\in\{0,1\}^n\}$, which means each encryption operator $X^xZ^z$ is carried out with the same probability $\frac{1}{2^{2n}}$. The attacker does not know which encryption operator is used, so the ciphertext is a completely mixed state $\frac{1}{2^{2n}}I$. This means QOTP has perfect security.

Compared with classical computation, quantum computation is usually more complex. This indicates that QFHE may be very different from classical FHE. Thus, an innovation is necessary for the research of QFHE. Here, we present a new definition about QHE as follows.

{\bf Definition 3 (QHE)}. A QHE scheme has a classical algorithm and three quantum algorithms: key generating algorithm, encryption algorithm, evaluate algorithm, and decryption algorithm.
\begin{itemize}
  \item
  Key generating algorithm is used to generate two keys: an encryption key $ek$ and a decryption key $dk$, where $ek$ is randomly selected and $dk$ is computed from $ek$ through a given algorithm.
  \item
  Encryption algorithm $\mathcal{E}:\rho=\mathcal{E}(ek,\sigma)$, where $\sigma$ is the plain data.
  \item
  Evaluate algorithm carries out a quantum computation $U$ on an encrypted data $\rho$, and an encrypted result $\rho'$ is obtained.
  \item
  Decryption algorithm $\mathcal{D}:\sigma'=\mathcal{D}(dk,\rho')$, where $\sigma'$ is the plain result. It is required that $\sigma'=U\sigma U^\dagger$.
\end{itemize}

It is worth to notice that, in the QHE cryptosystem, the encryption key is different from the decryption key, however, it is not a public key.

A QHE scheme is fully homomorphic, if it allows the evaluate algorithm to carry out any quantum computation on encrypted data.

\section{Schemes}
\subsection{Quantum homomorphic encryption scheme}
In this section, we present a construction of QHE based on a UQC. Here, suppose there exists a UQC $C_\mathcal{U}$ for certain family of quantum transformation $\mathcal{U}$, and the UQC $C_\mathcal{U}$ is consisted of the quantum gates from the set \{X, Y, Z, H, P, CNOT\}. According to the definition of UQC, $C_\mathcal{U}$ has two input interfaces, containing a $n$-qubit interface for data input and a $m$-qubit interface for encoding input.

Firstly, we introduce the key generating algorithm of the QHE scheme.

The encryption key $ek$ contains two $n$-bit random strings, which are chosen independently from the set $\{0,1\}^n$. To simplify the description later, each random string is appended $m$ bits $0\cdots 0$.The the encryption key is denoted as $ek=(x_0,z_0)$, where $x_0$ is a $n+m$-bit string $x_0(1)x_0(2)\ldots x_0(n+m)$ and $z_0$ is also a $n+m$-bit string $z_0(1)z_0(2)\ldots z_0(n+m)$. The bits $x_0(j),z_0(j) (n<j\leq n+m)$ are zeroes. So, the encryption key $ek$ has $2n$ random bits.

The algorithm of computing $dk$ from $ek=(x_0,z_0)$ is called key-updating algorithm, which is related to the UQC $C_\mathcal{U}$. In the QHE scheme, the construction of $C_\mathcal{U}$ is determined in advance. It consists of the quantum gates from the set \{X, Y, Z, H, P, CNOT\} by orderly combining several quantum gates. Assume there are total $k$ quantum gates in the circuit $C_\mathcal{U}$, then the key-updating algorithm can be realized with at most $k$ steps. Denote $(x_j,z_j),j=1,\ldots,k$ as the key obtained in the $j$th step of update. The key $(x_k,z_k)$ obtained finally is $dk$. The key-updating algorithm is the key part of our scheme, and will be proposed later.

Then, we introduce the encryption, evaluate and decryption algorithms of the QHE scheme.

According to QOTP, the $n$-qubit data $|d\rangle$ is encrypted using the key $ek=(x_0,z_0)$ as follows.
\begin{equation}
|d\rangle\rightarrow(\otimes_{w=1}^nX^{x_0(w)}Z^{z_0(w)})|d\rangle.
\end{equation}
The obtained ciphertext is input into the data interface of the UQC $C_\mathcal{U}$, and the $m$-qubit encoding $|e_U\rangle$ of a quantum transformation $U\in\mathcal{U}$ is input into the encoding interface. Then the UQC $C_\mathcal{U}$ would output the encrypted result as follows.
\begin{equation}
C_\mathcal{U}((\otimes_{w=1}^nX^{x_0(w)}Z^{z_0(w)})|d\rangle\otimes|e_U\rangle)=(U(\otimes_{w=1}^nX^{x_0(w)}Z^{z_0(w)})|d\rangle)\otimes|e_U\rangle.
\end{equation}
Later, we will show that the encrypted result $U(\otimes_{w=1}^nX^{x_0(w)}Z^{z_0(w)})|d\rangle$ has the form (up to an irrelevant global phase factor):
\begin{equation}
(\otimes_{w=1}^nX^{x_k(w)}Z^{z_k(w)})U|d\rangle,
\end{equation}
where $(x_k,z_k)$ is the decryption key $dk$.

In the decryption algorithm, the encrypted result is decrypted using the decryption key $dk=(x_k,z_k)$ as follows.
\begin{eqnarray}
&& U(\otimes_{w=1}^nX^{x_0(w)}Z^{z_0(w)})|d\rangle \nonumber\\
&\rightarrow & (\otimes_{w=1}^nX^{x_k(w)}Z^{z_k(w)})U(\otimes_{w=1}^nX^{x_0(w)}Z^{z_0(w)})|d\rangle \nonumber\\
&=& U|d\rangle.
\end{eqnarray}

Finally, we show how to update the encryption key $ek$ and obtain the decryption key $dk$.

According to our assumption, there are $k$ quantum gates in the UQC $C_\mathcal{U}$, and these quantum gates are labeled as $G_1, G_2, \ldots, G_k$ in accord with the orders of their execution. Each quantum gate is assumed to be one element of the set \{X, Y, Z, H, P, CNOT\}.
\begin{itemize}
\item If $G_j=X,Y$ or $Z$, the key does not change in the $j$th step of key-updating algorithm. Let
      \begin{equation}
      (x_j,z_j)=(x_{j-1},z_{j-1}),
      \end{equation}
      where the assignment ``$x_j=x_{j-1}$" means ``$x_j(i)=x_{j-1}(i),i=1,\cdots,n+m$".
\item If $G_j=H$ or $P$ (Assume it acts on the $w$th qubit, $w\in\{1,2,\ldots,n+m\}$), the $j$th step of key-updating only updates the $w$th bit of the key. Let $(x_j,z_j)=(x_{j-1},z_{j-1})$, and then changes the key as follows
    \begin{eqnarray}
    &&\mathrm{let}~(x_j(w),z_j(w))=(z_{j-1}(w),x_{j-1}(w)), \mathrm{if}~G_j=H;\\
    &&\mathrm{let}~(x_j(w),z_j(w))=(x_{j-1}(w),x_{j-1}(w)\oplus z_{j-1}(w)), \mathrm{if}~G_j=P;
    \end{eqnarray}
    where the notation ``$\oplus$" represents addition modular $2$.
\item If $G_j=$CNOT (Assume it acts on the $w$th and $w'$th qubits, where the $w$th qubit is the control and the $w'$th is the target. $w,w'\in\{1,2,\ldots,n+m\}$), the $j$th step of key-updating only updates the $w$th and $w'$th bits of the key. Let $(x_j,z_j)=(x_{j-1},z_{j-1})$, and then changes the key as follows
    \begin{eqnarray}
    &&\mathrm{let}~(x_j(w),z_j(w))=(x_{j-1}(w),z_{j-1}(w)\oplus z_{j-1}(w')),\\
    &&\mathrm{let}~(x_j(w'),z_j(w'))=(x_{j-1}(w)\oplus x_{j-1}(w'), z_{j-1}(w')).
    \end{eqnarray}
\end{itemize}

We have the following two results (see Ref. \cite{liang2013b}).
\begin{enumerate}
  \item
  The following relation holds for each quantum gate $G_j\in\{X,Y,Z,H,P\}$ (up to an irrelevant global phase factor).
  \begin{equation}\label{eq1}
  G_jX^{x_{j-1}(w)}Z^{z_{j-1}(w)}=X^{x_j(w)}Z^{z_j(w)}G_j.
  \end{equation}
  \item
  The following relation holds for the gate $G_j=$CNOT (up to an irrelevant global phase factor).
  \begin{eqnarray}\label{eq2}
  && G_j(X^{x_{j-1}(w)}Z^{z_{j-1}(w)}\otimes X^{x_{j-1}(w')}Z^{z_{j-1}(w')}) \nonumber\\
  &=& (X^{x_j(w)}Z^{z_j(w)}\otimes X^{x_j(w')}Z^{z_j(w')})G_j.
  \end{eqnarray}
\end{enumerate}
According to the above two relations, it can be deduced that (up to an irrelevant global phase factor)
\begin{eqnarray}
&& C_\mathcal{U}((\otimes_{w=1}^nX^{x_0(w)}Z^{z_0(w)})|d\rangle\otimes|e_U\rangle) \nonumber\\
&=& (\otimes_{w=1}^{n+m}X^{x_k(w)}Z^{z_k(w)})C_\mathcal{U}(|d\rangle\otimes|e_U\rangle) \nonumber\\
&=& (\otimes_{w=1}^nX^{x_k(w)}Z^{z_k(w)}U|d\rangle)\otimes(\otimes_{w=n+1}^{n+m}X^{x_k(w)}Z^{z_k(w)}|e_U\rangle).
\end{eqnarray}
So the cipher state $U(\otimes_{w=1}^nX^{x_0(w)}Z^{z_0(w)})|d\rangle$, which is obtained from the quantum computing on the encrypted data, can also be represented as this form $(\otimes_{w=1}^nX^{x_k(w)}Z^{z_k(w)})U|d\rangle$. Thus, the cipher state can be decrypted successfully with the decryption key $dk=(x_k,z_k)$.

Until now, we have presented a QHE scheme. The set of quantum gates \{X, Y, Z, H, P, CNOT\} is not universal for quantum computation \cite{nielsen2000,dupuis2010}, and cannot realize arbitrary quantum computation. When another quantum gate
$R=\left(
     \begin{array}{cc}
       1 & 0 \\
       0 & e^{i\pi/4} \\
     \end{array}
   \right)$
is added into the set, it becomes a universal set of quantum gates \{X, Y, Z, H, P, CNOT, R\}. Using the universal set of quantum gates, we can construct a UQC $C_\mathcal{U}$ for any quantum transformations. Then based on this UQC, a QFHE scheme can be constructed. The concrete description is shown in the next section.

\subsection{A quantum fully homomorphic encryption scheme}
Given a UQC $C_\mathcal{U}$ that consists of the quantum gates from the set \{X, Y, Z, H, P, CNOT, R\}, assume the UQC can realize any quantum transformation. Because the construction of the UQC $C_\mathcal{U}$ contains the $R$ gate, the QFHE scheme based on the UQC would be different from the QHE scheme in previous section. The difference is as follows. The encryption and decryption algorithms are the same, however, some new computation is added into the key-generating algorithm and evaluation algorithm. The details are shown in the following.

Firstly, it should be noticed that, interactive computation is necessary during the execution of evaluate algorithm. When the server performs the UQC on encrypted data, once a $R$ gate has been performed on a qubit (Assume the gate $G_j=R$ is performed on the $w$th qubit), then the server sends that qubit to the client; The client performs quantum operator $X^rZ^{r'}P^{x_{j-1}(w)}$ on the $w$th qubit, where $r,r'$ are two random bits selected by the client, and $x_{j-1}(w)$ is the $w$th bit of $x_{j-1}$ ($x_{j-1}$ is the client's key obtained in the $(j-1)$th step of key-updating algorithm).

Note that, the client's key-updating algorithm is synchronized with the execution of the UQC. According to the order of quantum gates in the UQC $C_\mathcal{U}$, the client updates his keys. Whenever encountering a $R$ gate, he waits for an interaction with the server. In the process of interaction, the client's key $x_{j-1}(w)$ would not be leaked, since he has selected two random bits $r,r'$ and carries out a QOTP encryption $X^rZ^{r'}$.

Then, we introduce the key-updating algorithm. In the process of key-updating, when $G_j=R$, let
\begin{equation}
(x_j(w),z_j(w))=(r\oplus x_{j-1}(w),r'\oplus x_{j-1}(w)\oplus z_{j-1}(w)),
\end{equation}
where $r,r'$ are two random bits selected by the client. Each time the values of $r,r'$ are selected independently.

It can be proved that $G_j=R$ satisfies the following relation (up to an irrelevant global phase factor).
\begin{equation}\label{eq3}
X^rZ^{r'}P^{x_{j-1}(w)}G_jX^{x_{j-1}(w)}Z^{z_{j-1}(w)}=X^{x_j(w)}Z^{z_j(w)}G_j.
\end{equation}
According  to the relations in Eqs.(\ref{eq1},\ref{eq2},\ref{eq3}), it can be deduced that, the output of the UQC (performed by the server) is
\begin{equation}
(\otimes_{w=1}^nX^{x_k(w)}Z^{z_k(w)}U|d\rangle)\otimes(\otimes_{w=n+1}^{n+m}X^{x_k(w)}Z^{z_k(w)}|e_U\rangle),
\end{equation}
where $\otimes_{w=1}^nX^{x_k(w)}Z^{z_k(w)}U|d\rangle$ is the encrypted result. The client can decrypt it successfully with the decryption key $dk=(x_k,z_k)$.

Until now, we have described a QFHE scheme in the two sections. The scheme is proposed based on a UQC, which is a combination of the quantum gates in the set \{X, Y, Z, H, P, CNOT, R\}. These gates can be used to construct a UQC which can realize any quantum transformation, thus our scheme has the property of fully homomorphic.

As an example, here introduce a construction of UQC, see Ref.\cite{liang2011}. This UQC can realize a class of basic quantum transformations (named near-trivial quantum transformation), and arbitrary quantum transformation $U$  can be implemented by a series of basic quantum transformation (for example, $U_1,\ldots,U_N$) in the class. Thus, based on the UQC, all these basic quantum transformations $U_1,\ldots,U_N$ can be realized orderly, and finally the desired quantum transformation $U$ is implemented. Note that, in order to implemented the transformation $U$, the UQC should be carried out for $N$ times.

Finally, it is worth to notice the following remarks.
\begin{itemize}
  \item The QFHE scheme contains an interactive process: if the $R$ gate occurs once in the UQC, an interaction is needed.
  \item The random bits $r,r'$ selected in the interaction will be used in the key-updating algorithm. Actually, these random bits can be chosen before the encryption/evaluation/decryption. If there are $n_R$ $R$ gates in the UQC, $2n_R$ random bits are necessary.
  \item The key-updating algorithm only depends on the construction of the UQC. So, given a UQC, the key-updating algorithm can be completed before the encryption/evaluation/decryption.
\end{itemize}

\subsection{Analysis}
We analyze the above QFHE scheme in this section. Because the QHE scheme is a limited case of the QFHE scheme, it is easy to analyze the QHE, and the analysis for the QFHE scheme is also suit for the QHE scheme.

The security is analyzed from two aspects: (1) the security of the keys, and (2) the security of plain data and result.

Let's firstly describe the whole process of the keys being used. The encryption key $ek$ is generated locally, and then be used in the QOTP encryption algorithm; Next, according to the UQC, the key is updated by the key-updating algorithm, and finally becomes to the decryption key $dk$, which is used in the decryption. Because of the property of QOTP, its ciphertext would not leak any information about the encryption key $ek$. In the whole process, the only step that may leak information is the key-updating process. During the interaction process of key-updating algorithm, when the server returns the $w$th qubit,  the client carries out a quantum operator $X^rZ^{r'}P^{x_{j-1}(w)}$ on it. Here, the random bits $r,r'$ are locally selected, and the quantum operator represents that a QOTP encryption is used after the execution of quantum gate $P^{x_{j-1}(w)}$. In this way, the bit $x_{j-1}(w)$ used in the interaction is protected and no information about the key is leaked. Thus, all the keys (the encryption/decryption keys and the updating keys) are perfectly secure.

Next, we analyze the security of plain data and result. The plain data is encrypted locally with the encryption key $ek$, and the encrypted data is input into the UQC; After the quantum computation is completed, the encrypted result is obtained; Finally, the plain result will be got through the QOTP decryption with the key $dk$. According to our scheme, the data $|d\rangle$ and final result $U|d\rangle$ remain in the encrypted states during the whole process of evaluation, such as $\otimes_{w=1}^nX^{x_0(w)}Z^{z_0(w)}|d\rangle$ and $\otimes_{w=1}^nX^{x_k(w)}Z^{z_k(w)}U|d\rangle$. Moreover, the intermediate results also remain in the encrypted states, such as $\otimes_{w=1}^nX^{x_j(w)}Z^{z_j(w)}|\varphi_j\rangle$ (Suppose the intermediate result in the $j$th step is $|\varphi_j\rangle$ when the plain data is input into the UQC). Because the keys are perfectly secure, according to the property of QOTP, the plain data and result are also perfectly secure.

Then, we analyze the computational complexity of the algorithms in the QFHE scheme, and the communication complexity in the process of interaction.

The encryption/decryption algorithms use the QOTP, so their computational complexity are $O(n)$. For the key generating algorithm, the key-updating algorithm computing $dk$ from $ek$ can be synchronized with the execution of UQC, and each step in the algorithm involves at most $3$ additive operations, so the computational complexity of key-updating algorithm is the same as that of the evaluate algorithm. Denote $|C_\mathcal{U}|$ as the circuit complexity (the number of quantum gates) of the UQC. Given a quantum computational task, if the UQC should be executed for $N$ times, then the evaluate algorithm has computational complexity $|C_\mathcal{U}|\cdot N$.

The interactive computation is only required in the evaluate algorithm of our scheme. Once a $R$ gate is executed, an interaction is needed: the server and the client send one qubit to each other. Denote $n_R$ as the number of the $R$ gates in the UQC, then the communication complexity is $2n_R\cdot N$.

\section{Discussions}
In Ref. \cite{liang2013a}, we had presented a kind of QFHE scheme. There are some differences between that scheme and the QFHE scheme in this paper. From the aspect of the keys, the former scheme uses the same encryption key and decryption key, but the decryption key in the latter scheme is different from the encryption key. However, the latter scheme is not a public-key homomorphic encryption scheme, since the encryption key cannot be public. From the aspect of evaluate algorithm, the former scheme does not have any interaction and the executor of the evaluate algorithm must know the encryption key, but it is not the case in the latter. This difference makes the latter more suitable for the secure delegated quantum computing between two parties.

Blind quantum computing is a kind of secure delegated quantum computing, in which the server performs quantum computation following the client's instructions and does not learn the client's algorithm. Compared with blind quantum computing, QFHE allows another kind of secure delegated quantum computing, in which the algorithm may be provided by the server or be determined by both the client and the server.

Our QFHE scheme is inspired by the TBQC scheme \cite{liang2013b}, and all the relations in Eqs.(\ref{eq1},\ref{eq2},\ref{eq3}) can be deduced directly from the relations in Ref. \cite{liang2013b}. After this work is completed, we find a similar research in Ref. \cite{fisher2014}, which proposed a scheme for quantum computing on encrypted data. In their scheme, the server must tell the client which algorithm (an algorithm means a series of quantum gates) is used to implement the quantum computation, then the client can update the key accordingly. However, in our scheme, the client's key-updating algorithm only depends on the construction of the UQC, and is irrelevant with the server's algorithm.

\section{Conclusions}
This paper proposes a new kind of QFHE scheme based on the UQC. QOTP is used in the encryption and decryption algorithm, and guarantees the perfect security. In the scheme, the encryption key and the decryption key are different. The encryption key is not public, and the decryption key can be computed from the encryption key. In addition, some interactive computation is necessary in the evaluate algorithm. The scheme is suitable for the delegated quantum computing between two parties: the client uses the selected key to encrypt the plain data, and sends the encrypted data to the server; then the server carries out quantum computation on the encrypted data without any knowledge about the key, and sends the encrypted result to the client; finally the client can decrypt it and obtain the plain result.

\end{document}